\documentclass[
preprint,
twocolumn,
5p,
number,
sort&compress
]{elsarticle}

\usepackage{amssymb,bbm}
\usepackage{graphicx}
\usepackage{hyperref}

\newcommand{\beq}{\begin{eqnarray}}
\newcommand{\eeq}{\end{eqnarray}}
\newcommand{\energ}{\mathcal{E}}

\begin{document}

\begin{frontmatter}
\title{ Renormalization group procedure for potential $-g/r^2$ }

\author[fuw,iu]{S. M. Dawid}
\corref{cor1}
\ead{sdawid@iu.edu}
\author[fuw]{R. Gonsior}
\author[fuw]{J. Kwapisz}
\author[fuw]{K. Serafin}
\author[impan,fuw]{M. Tobolski}
\author[fuw,yale]{S. D. G{\l}azek}

\cortext[cor1]{Corresponding author}
\address[fuw]{Institute of Theoretical Physics,
Department of Physics, University of Warsaw, ul. Pasteura 5, 02-093 Warsaw, Poland}
\address[iu]{Department of Physics,
Indiana University, Bloomington, IN 47405, USA}
\address[impan]{Institute of Mathematics, Polish Academy of Sciences, ul. {\'S}niadeckich 8,
00-656 Warsaw, Poland}
\address[yale]{Department of Physics, Yale University, New Haven, CT 06520, USA}

\begin{abstract}

Schr\"odinger equation with potential
$-g/r^2$ exhibits a limit cycle, described in the 
literature in a broad range of contexts using 
various regularizations of the singularity at 
$r=0$. Instead, we use the renormalization group 
transformation based on Gaussian elimination, 
from the Hamiltonian eigenvalue problem. of high 
momentum modes above a finite, floating cutoff 
scale. The procedure identifies a richer structure 
than the one we found in the literature. Namely, 
it directly yields an equation that determines 
the renormalized Hamiltonians as functions of 
the floating cutoff: solutions to this equation 
exhibit, in addition to the limit-cycle, also the 
asymptotic-freedom, triviality, and fixed-point 
behaviors, the latter in vicinity of infinitely 
many separate pairs of fixed points in different 
partial waves for different values of $g$.
\end{abstract}


\date{18 August 2017}

\end{frontmatter}

\section{Introduction}
\label{sec:I}
Eigenvalue problem for the one-particle Hamiltonian 
\beq
\label{eq:hamiltonian-poczatkowy}
H = \frac{ \vec p\,^2}{2m} - \frac{g}{\vec r \, ^2} \  ,
\eeq
provides a well-known example of a singular
Schr\"odinger equation. For positive and
sufficiently large coupling constant $g$, the
overwhelmingly attractive potential causes
instability, which limits application of
Eq.~(\ref{eq:hamiltonian-poczatkowy}) in 
physics~\cite{Case51}.

The situation is changed when one regularizes the
potential and introduces corrections on the basis
of demanding that predictions for observables do
not depend on the regularization. As a result, the
corrected interaction exhibits cyclic behavior as a
function of the regularization cutoff parameter.
The cycle is associated with an infinite set of 
bound states whose binding energies form a geometric
sequence converging on zero~\cite{Case51,Beane01,
BalCoon03,CoHol02,BraPhil04,HammSwin06,HammerHiga,
LoKol08,Pavon08}.

Regularization of potential $-g/r^2$ in the 
position representation is proposed in various 
ways~\cite{BraPhil04,Beane01,BalCoon03,KaLeeSonSte09}. 
For instance, Braaten and Phillips~\cite{BraPhil04} 
cut off the potential at some small radius and they 
introduce an additional potential that acts only in 
a spherical $\delta$-shell of an infinitesimally
smaller radius. They solve the resulting $s$-wave
eigenvalue problem and find that one can obtain
cutoff independent eigenvalues by making the coupling 
constant in the additional term a log-periodic function 
of the cutoff radius.

In the momentum representation, the regularization
is formulated differently. For example, in the
space of $s$-wave states, Hammer and
Swingle~\cite{HammSwin06} introduce an ultraviolet
cutoff $\Lambda$ that limits the particle momentum
from above. They also add a counter potential with
a new coupling constant, $H(\Lambda)$. They
determine $H(\Lambda)$ as a function of $\Lambda$
by demanding that the bound-state solution with
zero binding energy does not depend on $\Lambda$.
Knowing the function $H(\Lambda)$, they find a
differential equation that it satisfies. In higher
partial waves, Long and van Kolck~\cite{LoKol08}
discuss the potential $-g/r^2$ with added contact
terms, in the spirit of constructing effective 
theories. They arrive at cutoff-independent solutions
for observables by specifying cycling coefficients
in front of one contact term per singular partial
wave.

The approach quoted above has successfully amended
Eq.~(\ref{eq:hamiltonian-poczatkowy}) with examples 
of regularization and corresponding correction terms 
that are justified {\it a posteriori}. The ultimate 
justification relies on the fact that different
regularizations with different corrections lead to
the same results for observables~\cite{Cao:1993gpm,
Cao:1999pw}.

In this paper, the Hamiltonian of
Eq.~(\ref{eq:hamiltonian-poczatkowy}) is handled
using the Wilsonian type of renormalization group
procedure~\cite{Wil65,Wil70}. It is the same type
of procedure that originally produced the concept 
of a renormalization group limit
cycle~\cite{Wilson:1970ag}. Instead of first
calculating observables in a regularized theory,
 guessing an ansatz for the correction term,
and subsequently checking if such a theory can
give regularization independent results for the
observables, we begin by {\it calculating} the
required counterterm structure in the presence 
of regularization. This is done using a
renormalization group transformation (RGT) of the
Hamiltonian itself~\cite{Wil65,Wil70}, {\it i.e.},
prior to seeking solutions for observables.
In the process, a family of finite, effective
Hamiltonians is obtained, using the calculated 
counterterm structure. Solutions for observables 
are sought first in the resulting finite, 
scale-dependent effective theories that {\it a 
priori} do not depend on regularization. 

Our procedure leads to the limit cycle in terms 
of solutions to a simple renormalization group 
equation that evolves a Hamiltonian in the sense 
of Refs.~\cite{Wil65,Wil70}. The procedure also 
allows us to identify a whole range of renormalization 
group behaviors, which characterize the Hamiltonian 
of Eq.~(\ref{eq:hamiltonian-poczatkowy}), in addition 
to the limit cycle. These other behaviors, to the best
of our knowledge, are not fully identified in the literature.
Namely, we exhibit behaviors of the asymptotic-freedom,
triviality and fixed-point type. The latter has been 
found before in the case of $s$-waves, using the functional 
renormalization group technique that revealed a collision 
of two fixed points at the critical value of the $s$-wave 
coupling constant~\cite{Moroz:2009nm}. We identify fixed 
points in all partial waves, including their behavior near 
the corresponding infinitely many critical values of the 
coupling constant. These behaviors appear in a pattern of 
interest for studies of scaling symmetry and its breakdown 
in complex theories, see below. Hence, the simplicity 
and familiarity of Eq.~(\ref{eq:hamiltonian-poczatkowy}) 
are its assets rather than drawbacks. Further 
simplification is taken advantage of in this paper 
by carrying out the RGT in the differential steps 
that are analogous to the discrete ones outlined 
in~\cite{SDGWil02,SDGWilErr} on another example 
of a Hamiltonian with a limit cycle. The issue of 
dependence of effective theories on the magnitude 
of eigenvalues they aim to describe, is only briefly
explained at the end of the paper.

Effective potentials of type $1/r^2$ were
successfully employed in three-body dynamics,
where a series of bound states generically
appears, under the name of Efimov
effect~\cite{Thomas:1935zz,Efimov1970,Efimov1973,Bedaque},
whenever in the corresponding two-body dynamics
the ratio of effective interaction range to
scattering length is very small, ultimately
requiring precise treatment when this ratio
approaches zero~\cite{Mohr2006}. It is worth
pointing out that a potential of the type $1/r^2$
appears also in interactions of a point charge
with a dipole, causing violation of scaling
symmetry~\cite{CamEpe00,CamEpe01}. The pattern of
breaking scale invariance when the coupling
constant $g$ in
Eq.~(\ref{eq:hamiltonian-poczatkowy}) increases
above a critical value is of special interest in
many areas ``from molecular to black-hole
physics''~\cite{Camblong2003}, and in statistical
mechanics~\cite{Kosterlitz,Kolomeisky2}. In
Ref.~\cite{KaLeeSonSte09}, the analogous pattern
is found helpful in discussing theories that
exhibit the Berezinsky-Kosterlitz-Thouless phase
transition. It is said to have many ``parallels in
the AdS/CFT correspondence''~\cite{Maldacena}.
Patterns of conformal symmetry breaking, including
breaking of scale invariance in
Hamiltonians~\cite{Alfaro}, are invoked in
light-front holographic description of
hadrons~\cite{Brodsky} and in studies of conformal
windows in technicolor gauge
theories~\cite{Dietrich}, the latter hoped to help
in understanding the origin of the Standard
Model~\cite{BanksZaks,Appelquist}. 
Equation~(\ref{eq:hamiltonian-poczatkowy}) is thus
of broad interest as a rich but simple example of 
scaling-symmetry breaking and dimensional 
transmutation~\cite{ColemanWeinberg} in the Wilsonian 
renormalization group procedure for Hamiltonians.

\section{ Renormalization group transformation }
\label{sec:II}

The renormalization group procedure for an
incompletely defined Hamiltonian, such as the one
in Eq.~(\ref{eq:hamiltonian-poczatkowy}), starts
with regularizing it as an operator in a scheme
called the ``triangle of renormalization''~\cite{Wilsonetal}. 
We introduce a cutoff $\Delta$ and thus define $H_\Delta$, 
from which we evaluate the equivalent effective
Hamiltonians $H_\Lambda$ with cutoffs $\Lambda \ll
\Delta$. The RGT is involved in the process of 
this evaluation.

In the limit $\Delta \to \infty$, we demand that 
matrix elements of $H_\Lambda$ in the subspace of 
states limited by finite $\Lambda$ do not depend 
on $\Delta$. This condition implies a large, if 
not infinite set of constraints. It allows us to 
determine the structure of counterterms needed in 
$H_\Delta$, up to finite alterations in their 
parameters. The computation of counterterms is 
done in a sequence of successive approximations, 
which improves the counterterms put in $H_\Delta$ 
until all matrix elements in $H_\Lambda$ in the 
subspace of states limited by $\Lambda$ become 
independent of $\Delta$. This finite cutoff 
becomes a running parameter which enables one 
to identify how the theoretical description of 
observable phenomena depends on the range of 
scales of degrees of freedom one uses for
constructing solutions of the theory.

Once the counterterms in $H_\Delta$ are such that 
all matrix elements of $H_\Lambda$ have well-defined 
limits for $\Delta/\Lambda \to \infty$, the
eigenvalues of $H_\Lambda$ cannot depend on $\Delta$. 
At the same time, eigenvalues of the effective 
Hamiltonians $H_\Lambda$ that are much smaller than 
$\Lambda$ cannot depend on $\Lambda$, because this cutoff
 is merely a mathematical boundary
between the implicit degrees of freedom above and
explicit degrees of freedom below $\Lambda$. Computation 
of $H_\Lambda$ that is equivalent to $H_\Delta$ is 
carried out in a sequence of discrete, or infinitesimal 
RGTs. 


We carry out a sequence of the RGTs in momentum 
representation. The stationary Schr\"odinger 
equation for Hamiltonian of Eq.~(\ref{eq:hamiltonian-poczatkowy}),
\beq
\label{momentum space eq final} 
\frac{\vec{p}^{\ 2}}{2m} \phi(\vec{p}) - {g \over 4\pi} 
\int d^3q \ \frac{ \phi(\vec{q}) }{ |\vec{p} -
\vec{q}| }  = E \phi(\vec{p}).
\eeq
is written in terms of angular and radial 
variables, $\phi(\vec{p}) = \sum_{lm} 
\psi_l(p) Y_{lm}(\Omega_p)$, where $p = 
|\vec p \, |$. Taking advantage of rotational 
symmetry of Eq.~(\ref{momentum space eq final}), 
one obtains 
\beq
\label{eq:pedrad1}
p^2 \psi_l(p) + \int\limits_0^\infty \! \! dq~q^2 \ V_l(p,q)
\ \psi_l(q) =  \energ \psi_l(p) ,
\eeq
where the eigenvalue $\energ = 2mE$, the potential 
\beq
\label{eq:potential1}
V_l(p,q) = -\frac{\alpha}{2l + 1}~ 
\left[\frac{ \theta(p-q)~q\,^l}{ p\,^{l+1}} + 
\frac{ \theta(q-p)~ p\,^l}{q\,^{l+1}}\right] ,
\eeq
the dimensionless coupling constant  $\alpha 
= 2 mg$ and the symbol $\theta$ denotes the 
Heaviside step function. Our regulated 
eigenvalue problem is defined by limiting 
the range of momenta $p$ and $q$ by a large 
cutoff parameter $\Delta$.

Elimination of high momentum modes proceeds 
by infinitesimal steps such as from $\Delta$ 
to $\Delta - d\Delta$,
\beq
\label{eq:reduction1}
p^2 \psi_l(p)\!\! & + &\!\! \int\limits_0^{\Delta - d\Delta} \! dq \ q^2 \
V_l(p,q) \ \psi_l(q) \nonumber \\
\! & - &\! \frac{\alpha}{2l+1} \, \frac{p\,^l}{\Delta^{l-1}} \, 
\psi_l(\Delta) \ d\Delta = \energ \ \psi_l(p) \ .
\eeq
The value of $\psi_l(\Delta)$ is expressed in terms of 
values of $\psi_l(p)$ with $p < \Delta - d\Delta$ by 
setting $p=\Delta$ in Eq.~(\ref{eq:reduction1}).
For eigenvalues $\energ$ much smaller than
$\Delta^2$, and neglecting terms that lead to 
quantities of order $d\Delta^2$ or smaller in 
Eq.~(\ref{eq:reduction1}), we have
\beq
\label{eq:wavefunct}
\psi_l(\Delta) = \frac{\alpha}{(2l+1) \Delta^2} 
\int\limits_0^{\Delta-d\Delta} \! dq \ q^2 \
\frac{q\,^l}{\Delta^{l+1}} \ \psi_l(q)\:.
\eeq
Hence, treating $\Delta - d\Delta$ as a new cutoff
$\Lambda$, we obtain
\beq
\label{eq:pedrad2}
p^2 \ \psi_l(p) \!\!\!\!\!
& + & \!\!\!\!\!
\int\limits_0^\Lambda \! dq  \ q^2 
\left[ V_l(p,q) + \gamma_\Lambda q\,^l p\,^l \right] \ \psi_l(q)
\nonumber \\
& = & \!\!\!
\energ \ \psi_l(p) \ , \\
\gamma_\Lambda \!\!\! & = & \!\!\! - 
\frac{\alpha^2 d\Delta}{(2l+1)^2 \Delta^{2l+2}} \ .
\eeq
Further reduction of the cutoff from $\Lambda$ to $\Lambda - d\Lambda$ 
produces
\beq
\label{eq:reduction2}
p^2 \ \psi_l(p) \!\!\!
& + & \!\!\!\!\!\!\!
\int\limits_0^{\Lambda - d\Lambda} \! dq \ q^2 
\left[ V_l(p,q) + \gamma_{\Lambda - d\Lambda} q\,^l p\,^l \right]  \psi_l(q)
\nonumber \\
& = &\!\!
\energ \ \psi_l(p) \ ,  
\eeq
where
\beq
\label{eq:rge1}
\gamma_{\Lambda-d\Lambda} = 
\gamma_\Lambda - \left[ \gamma_\Lambda \Lambda^l - 
\frac{\alpha}{(2l+1)\Lambda^{l+1}} \right]^2
d\Lambda \ .
\eeq
Now no new potential term appears. Only the
coupling constant $\gamma_\Lambda$ changes to
$\gamma_{\Lambda-d\Lambda}$. After repeating 
such steps a large number of times, one finds 
that, for as long as $\Lambda^2 \gg \energ$, and
for every eigenvalue that is much smaller than 
$\Lambda^2$, the only change in the interaction 
is in the constant $\gamma$. Therefore, $\gamma$  
as a function of $\Lambda$ obeys the Ricatti 
equation
\beq
\label{eq:rge}
\frac{d \gamma}{d \Lambda} 
= 
\left[ \gamma \Lambda^l - \frac{ \alpha}{(2l+1)
\Lambda^{l+1}} \right]^2 \ .
\eeq
The coupling constants $\gamma$ for different 
values of the angular momentum quantum number 
$l$ obey different equations.

A change in $\gamma$ causes changes in all matrix 
elements of the Hamiltonian in the corresponding 
partial-wave subspace. However, all these changes 
are described in terms of variation of one number 
in front of a term of precisely determined structure. 
Thus, the RGT explains why satisfying just one 
eigenvalue condition per partial wave can make 
all other eigenvalues well below the cutoff not 
depend on the cutoff. 

The counterterm in the Hamiltonian $H_\Delta$ 
is obtained by sending $\Lambda$ to the ultimate 
limit of $\Delta$ that is considered equivalent 
to infinity in comparison to all finite parameters
and results of the theory. The counterterm is the 
term needed to reset the initially unknown coupling 
constant $\gamma_\Delta$ to the value implied by 
the right value of $\gamma_\Lambda$ in $H_\Lambda$ 
at some finite $\Lambda$, see below. This implication
is realized in terms of solutions of 
Eq.~(\ref{eq:rge}). As a result, all matrix elements 
of $H_\Lambda$ are independent of $\Delta$. 

The right value of $\gamma_\Lambda$ can be found
by reproducing one physical quantity associated
with the corresponding value of angular momentum 
quantum number $l$. The freedom of choosing the 
right value of $\gamma_\Lambda$ corresponds to the 
freedom of choosing the initial condition for it 
at $\Lambda = \Delta$ while solving Eq.~(\ref{eq:rge}). 
Such initial condition is missing in the incomplete 
definition of the Hamiltonian in 
Eq.~(\ref{eq:hamiltonian-poczatkowy}).

Instead of using an initial condition for $\gamma$
at $\Lambda = \Delta$, one can use a condition
that at some finite $\Lambda_0$, the coupling
constant must have a finite value $\gamma_0$ in
order to produce a physically specified eigenvalue
$\energ_0$. This is how one arrives at the
coupling constant $\gamma$ that is a function of
the ratio $\Lambda/\energ_0$ and which takes the
value $\gamma_0$ at $\Lambda_0$, breaking scaling
symmetry of Eq.~(\ref{eq:hamiltonian-poczatkowy})
in agreement with the concept of dimensional
transmutation~\cite{ColemanWeinberg}. 

The reason for the RGT to yield so simply a family
of renormalized Hamiltonians for
Eq.~(\ref{eq:hamiltonian-poczatkowy}), is the
simplicity of the latter. In general, many
operator structures may be involved and one needs
to approximate solutions in a series of successive
approximations.

A comment is in order concerning the fact that
Eq.~(\ref{eq:rge}), when considered in isolation, 
is resembled by the equations found in the 
effective-theory approach~\cite{HammSwin06,LoKol08}
mentioned in Sec.~\ref{sec:I}. The conceptual difference 
stems from the RGT, whose application here to 
the best of our knowledge is new and leads to new 
findings described below in Sec.~\ref{sec:solutions}. 
The effective-theory approach proposes that some 
specific ansatz contact terms are sufficient for 
removing regularization dependence from low-energy 
observables. The proposal is checked by solving 
the regularized theory with the ansatz terms and 
showing that it is indeed the case, if the ansatz 
terms as functions of regularization satisfy the 
equation in question. The RGT shows instead that 
Eq.~(\ref{eq:rge}) is valid for Hamiltonians with finite 
cutoffs, in principle independently of regularization 
and without invoking any ansatz. More generally, 
when one uses solutions to a regularized theory 
in the no cutoff limit, the verification of an 
ansatz requires more complex computation than the 
RGT, because the latter is carried out without 
solving a theory simultaneously at all scales,
{\it i.e.}, for a cutoff going to infinity. In 
view of the relative complexity of searching for 
such solutions, it is encouraging that in the 
case of Eq.~(\ref{eq:hamiltonian-poczatkowy}) 
one can verify the ansatz terms not only for the 
$s$-waves~\cite{HammSwin06} but also for higher 
partial waves~\cite{LoKol08} and establish their 
limit-cycle behavior. However, the RGT Eq.~(\ref{eq:rge}) 
is simply obtained as valid without any guessing
and it also describes the renormalization group 
behaviors that appear in addition to the limit 
cycle. The latter are identified in Sec.~\ref{sec:solutions}
below. In realistic theories, exact solutions with 
diverging cutoffs appear to be out of reach while 
the RGT studies still appear doable and patterns 
like the ones found in Sec.~\ref{sec:solutions} 
may be useful in understanding complex theories.

\section{ Solutions for renormalized Hamiltonians }
\label{sec:solutions}

Writing $\gamma_\Lambda = f/\Lambda^{2L}$ with
$L = l + 1/2$, one turns Eq.~(\ref{eq:rge}) into
a $\beta$-function of dimensionless coupling 
constant $f$, 
\beq
\label{rightone}
\Lambda \, \frac{\partial f}{\partial \Lambda} 
=
\beta(f) 
= (f-A)^2 + B^2 \ ,
\eeq
where
\beq
\label{parAB}
A & = & \alpha/(2L)  - L , 
~~B = \sqrt{\alpha - L^2 } \ . 
\eeq
For real $B\neq0$, or $\sqrt{\alpha} > L$, 
the solution is
\beq
\label{eq:riccati solution}
f = \left[B \tan \left( \varphi  + B  
\ln \frac{\Lambda}{\Lambda_0}  \  \right) + A \right] \ ,
\eeq
where
\beq
\varphi = \arctan \left( \frac{f_0 - A}{B}  \right) 
\eeq
secures the desired $l$-dependent value of $f_0 = 
f(\Lambda_0)$, established by using $H_{\Lambda_0}$ 
to reproduce one physical energy level per partial 
wave.

Adjustment of $f_0$ using $H_{\Lambda_0}$ to
produce some desired bound-state eigenvalue
$\energ \ll \Lambda_0^2$, produces infinitely many
smaller in magnitude bound-state eigenvalues in
the form of a cutoff-independent geometric
sequence converging on zero with quotient $r =
\exp(-2\pi/B)$. This can be seen using
Eq.~(\ref{eq:pedrad2}), where introduction of
dimensionless variables $p/\Lambda$ and
$q/\Lambda$ instead of $p$ and $q$ shows that the
eigenvalues are proportional to $\Lambda^2$ with
coefficients depending on the numbers $\alpha$ and
$f$. Since $\alpha$ is constant and $f$ is
log-periodic in $\Lambda$, the change of
$\Lambda^2$ by the discrete periodicity factor $r$
does not change the proportionality coefficient
between the eigenvalue and $\Lambda^2$. At the
same time, the spectrum of eigenvalues $\energ \ll
\Lambda^2$ does not depend on the value of
$\Lambda$. Thus, existence of an eigenvalue
$\energ$ implies existence of an eigenvalue $r
\energ$. This reasoning establishes the cycle.
Namely, if the coupling constant $\alpha >
\alpha_{\, lc} = L^2$, the limit cycles
appear in the partial waves up to $l$. There is 
no limiting value of $f$ in $H_\infty$ in these
waves, because $f$ oscillates forever when the
cutoff $\Lambda$ is increased. In summary, the 
apparent continuous scaling symmetry of 
Eq.~(\ref{eq:hamiltonian-poczatkowy}) is 
broken to a discrete scaling symmetry with 
the factor $r$.

When $B$ is purely imaginary, which occurs when
$\alpha < \alpha_{\, lc}$, implying $L >
\sqrt{\alpha}$, the behavior of the coupling
constant $f$ is governed by two fixed points of
Eq.~(\ref{rightone}), $f_\pm = A \pm |B| = -(L/2)(1
\mp \sqrt{ 1 - \alpha/L^2})^2 < 0$. The one that
is more negative, $f_-$, is repulsive and $f_+$
is attractive when we lower $\Lambda$. Therefore, 
$f$ decreases to $f_+$ from any value above it,
moves in the direction from repulsive to attractive
between $f_-$ and $f_+$, or decreases further down 
from any value below $f_-$. When the direction of 
change in the floating cutoff is reversed, one 
obtains the Landau pole starting anywhere above 
$f_+$, transition toward $f_-$ from anywhere below 
$f_+$ but above $f_-$, and increase toward the latter 
from anywhere below it.

\begin{figure}[t!]
\begin{center}
\includegraphics{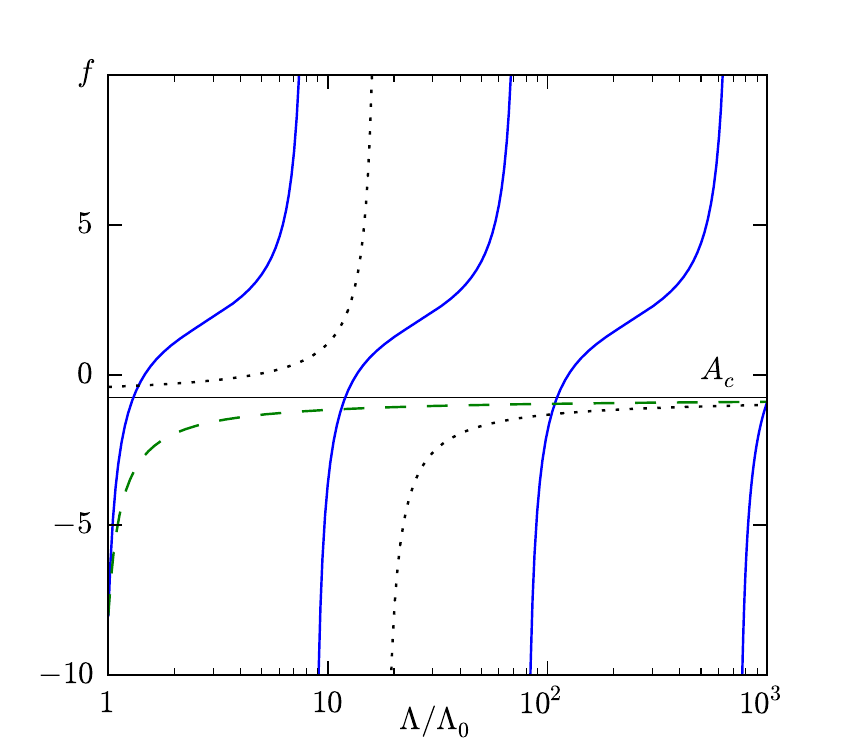} 
\caption{ Examples of dependence of the counterterm
coupling constant $f$ for two lowest angular momentum 
quantum numbers $l$ and the coupling constant $g$ 
in Eq.~(\ref{eq:hamiltonian-poczatkowy}) equal $9/(8m)$, 
which is a critical value for $l=1$, with the corresponding
value of $A_c = -0.75$ indicated by a horizontal line. For
$l=0$, $f$ exhibits limit-cycle (solid line). For $l=1$,
$f$ approaches $A_c$ in a logarithmic fashion that
characterizes approach to zero in asymptotic-freedom
(dashed line). In both cases $f(\Lambda_0) = -8.0$. 
Setting $f(\Lambda_0) = -0.4$, which is above $A_c$, 
one obtains an example of the Landau pole (dotted line).}
\label{fig:1}
\end{center}
\end{figure}

When $\alpha = \alpha_{lc}$, in which case $l=0$
corresponds to the well-known critical value 
$\alpha_{0c} = 1/4$, one obtains 
\beq
\label{AFLP}
f-A_c = {f_0 - A_c \over 1 - (f_0 - A_c) \ln {\Lambda
\over \Lambda_0 } } \ ,
\eeq
where, for the critical value of $l$ for this
$\alpha$, $A_c=f_+=f_- = - L/2$. Thus, considering 
only  $\Lambda$ that increases above some $\Lambda_0$, 
one obtains the renormalized Hamiltonians with $f$ 
that approaches the fixed point $f = A_c$ in a 
logarithmic way characteristic of the asymptotic-freedom 
when one starts from $f_0 < A_c$, or deviates from 
$A_c$ in the fashion of triviality, associated with 
the Landau pole, when $f_0 > A_c$. This result extends 
the previously noted relationship between renormalization 
group limit-cycle and asymptotic-freedom behaviors~\cite{sdgET} 
to the case of $-g/r^2$ potential in the Schr\"odinger 
equation, since $\alpha \to \alpha_{lc}$ from above
implies $B \to 0$ and the cycle period growing to
infinity. On the other hand, when $\alpha$ approaches
$\alpha_{lc}$ from below, the distance between the two
fixed points $f_-$ and $f_+$ for the critical value of
$l$ shrinks to zero. 

\begin{figure}[t]
\begin{center}
\includegraphics{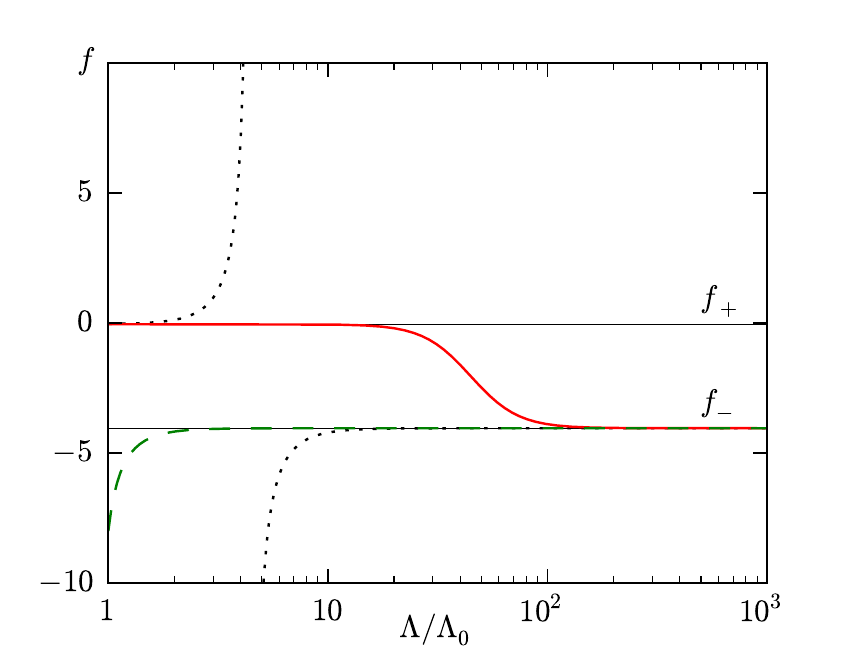} 
\caption{ Examples of dependence of the counterterm
coupling constant $f$ for $l=2$ and the same coupling
constant $g = 9/(8m)$ as in Fig.~\ref{fig:1}, critical 
for $l=1$. The two fixed points are $f_+ = - 0.05$
and $f_- = -4.05$. For $f(\Lambda_0) = f_+ - 10^{-6}$, 
$f$ flows toward $f_-$ (solid line). For $f(\Lambda_0) 
= f_+ + 10^{-2}$, $f$ develops the Landau pole (dotted 
line). For $f(\Lambda_0) = -8.0 < f_-$, $f$ increases to 
$f_-$ (dashed line).
}
\label{fig:2}
\end{center}
\end{figure}

\begin{figure}[t]
\begin{center}
\includegraphics{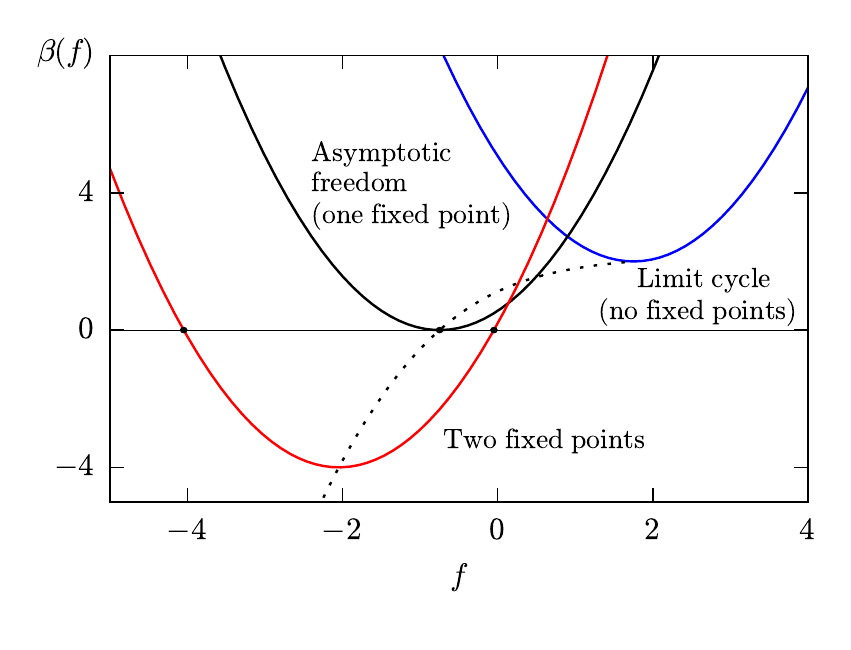} 
\caption{ Beta-functions of the coupling constants $f$
in partial waves $l=0,1,2$, corresponding to the
renormalization group behaviors illustrated in
Figs.~\ref{fig:1} and \ref{fig:2}. The dotted line 
is the curve on which lie the minima of all beta-functions, 
parameterized by $l$, see Eq.~(\ref{parAB}). As $l$
increases, and $\alpha$ is fixed, the minimum goes
down the line. When $\alpha$ is increased, the whole
line shifts up and right (and deforms) allowing more 
partial waves to enter into the limit cycle regime. 
This illustrates how transition from fixed points 
through asymptotic freedom to limit cycle occurs.}
\label{fig:3}
\end{center}
\end{figure}

Examples of behavior of the constant $f$ as 
a function of the cutoff $\Lambda$ are shown 
in Figs.~\ref{fig:1} and \ref{fig:2}. The figures 
display the flow of the couplings for three different 
angular momenta, $l = 0, 1, 2$. The coupling constant 
$\alpha$ is set to the critical value for $l=1$. 
It is visible that one and the same Hamiltonian 
exhibits various renormalization group behaviors 
in different partial waves. Figure~\ref{fig:3} 
presents the beta-functions for the same cases. 
The asymptotic freedom case appears in the vicinity 
of one fixed point that can be thought to result 
from a collision of two fixed-points, in analogy 
to the $s$-wave analysis in Ref.~\cite{Moroz:2009nm}.
Figure~\ref{fig:3} also shows that the asymptotic-freedom 
case can be seen as resulting from the limit cycle 
when the minimum of beta-function reaches zero and 
the period of the cycle grows to infinity. In each 
case one can adjust the corresponding value of $f_0$ 
to a desired binding energy or a phase shift at a 
selected energy. However, a complete description of 
various types of bound-state and scattering observables 
for different choices of finite parameters in the 
renormalized Hamiltonians is beyond the scope of 
this paper.

\section{Conclusion}
\label{sec:III}

Simplicity of the renormalization procedure 
for $-g/r^2$ potential in the Schr\"odinger 
equation, carried out here using the RGT of 
Wilsonian type, results from neglecting the 
ratio of eigenvalues $E$ to the floating cutoff 
$\Lambda$. Inclusion of this ratio in the RGT 
can be carried out according to the steps 
described in Ref.~\cite{SDGWil04}. For example, 
one can employ an expansion in powers of $E/\Lambda$ 
and establish corrections needed in $H_\Delta$ 
to accelerate convergence of the family $H_\Lambda$ 
on the limit cycle when $\Lambda$ is lowered. When 
in practice the ratio $E/\Lambda$ is not sufficiently 
small and the expansion in its powers is not efficient, 
one can switch to the similarity renormalization 
group procedure~\cite{SRG1,SRG2}. In principle, the 
similarity procedure avoids dependence on the eigenvalues 
and can be employed as in Refs.~\cite{sdgET, NiemannHammer}.\\~\\
\textbf{Acknowledgements}\\

Authors thank the Faculty of Physics at the University of
Warsaw for support of the course {\it Introduction to Renormalization}
within which the research reported here was carried out.


\bibliography{mybibfile-20170719}{}

\end{document}